\documentclass[mathpazo]{cicp}

\let \nn  \nonumber

\def\be{\begin{equation}}\def\ee{\end{equation}}
\def\bea{\begin{eqnarray}}\def\eea{\end{eqnarray}}
\def\bse{\begin{subequations}}\def\ese{\end{subequations}}
\newcommand{\BE}[1]{\begin{equation}\label{#1}}
\newcommand{\BEA}[1]{\begin{eqnarray}\label{#1}}
\newcommand{\BSE}[1]{\begin{subequations}\label{#1}}

\usepackage{algorithm}
\usepackage{algorithmic}
\let \nn  \nonumber
\usepackage{pstricks}
\usepackage{pst-node}
\usepackage[ansinew]{inputenc}
\usepackage{amssymb,amsmath}

\def\BSE{\begin{subequations}}\def\ESE{\end{subequations}}
\let \= \equiv

\def\be{\begin{equation}}       \def\ba{\begin{array}}

\def\ee{\end{equation}}         \def\ea{\end{array}}

\def\bea {\begin{eqnarray}}      \def\eea {\end{eqnarray}}

\def\bean{\begin{eqnarray*}}    \def\eean{\end{eqnarray*}}

\def\const {\mathop{\rm const}\nolimits}

\def\RA {\ \Rightarrow\ }

\def\<{\langle} \def\({\left(}  \def\>{\rangle} \def\){\right)}

\begin{document}
\title[short title]{Generalized Manley--Rowe constants}
\title{A constructive method for computing generalized Manley--Rowe constants of motion}

\author[E. Kartashova and L. Tec]{Elena Kartashova\affil{1}\comma\corrauth and  Loredana Tec\affil{2}}
\address{\affilnum{1}\ IFA, J. Kepler University, Linz, Austria\\
\affilnum{2}\ RISC, J. Kepler University, Linz, Austria}
\emails{{\tt Elena.Kartaschova@jku.at} (E.~Kartashova), Loredana.Tec@jku.at (L. Tec)}


\begin{abstract}
The Manley--Rowe constants  of motion (MRC) are conservation laws written out for a dynamical system describing the time evolution of the amplitudes in resonant triad. In this paper we extend the concept of MRC to resonance clusters of any form yielding generalized Manley--Rowe constants (gMRC) and give a constructive method how to compute them. We also give details of a \emph{Mathematica} implementation of this method. While MRC provide integrability of the underlying dynamical system, gMRC generally do not but may be used for qualitative and numerical study of dynamical systems describing generic resonance clusters.
\end{abstract}


\maketitle

\section{Introduction}

Weakly nonlinear wave systems showing resonances can be found in a great variety of physical systems, among them surface water waves, atmospheric planetary waves, plasma drift waves, etc.
In Hamiltonian formulation  the  equation
 of motion in Fourier space can be written out as
\be i\,\dot B_{\mathbf{k}} = \partial {\mathcal{H}}/\partial B_{\mathbf{k}}^*,
\label{HamiltonianEquationOfMotion} \ee
 where $B_{\mathbf{k}}$
is  the amplitude of the Fourier mode corresponding to
wavevector $\mathbf{k}$ and the Hamiltonian $\mathcal{H}$ is represented
as an expansion in powers of terms ${\mathcal{H}}_j$, each representing all products of $j$ amplitudes $B_{\mathbf{k}}$:
\be \label{Hexp}
\mathcal{H} = \mathcal{H}_2 + \mathcal{H}_{int}\,, \ \
 \mathcal{H}_{int}=\mathcal{H}_3+ \mathcal{H}_4 +  \mathcal{H}_5 + \dots\,.
 \ee
The quadratic
Hamiltonian $ \mathcal{H}_{2}$
describes  linear motion and
the interaction Hamiltonian $ \mathcal{H}_{int}$ describes nonlinear  interaction of waves.

 If the cubic Hamiltonian ${\mathcal{H}}_3 \ne 0$, $\mathcal{H}_{int} \approx \mathcal{H}_3$,  three-wave interaction is dominant and the main contribution to the nonlinear evolution comes from triads of waves each satisfying the resonance conditions%
 \be
\label{res}
\omega (\mathbf{k}_1) + \omega (\mathbf{k}_2)- \omega (\mathbf{k}_{3}) = 0,\ \
\mathbf{k}_1 + \mathbf{k}_2 -\mathbf{k}_{3} = 0.
\ee %
 The corresponding
dynamical system describing time evolution of the amplitudes $B_j$ of a triad reads
\be\label{ch4:complexB0}
\dot{B}_1=  Z B_2^*B_3,\,
\dot{B}_2=  Z B_1^* B_3, \, \dot{B}_3=  - Z B_1 B_2,\ee
the interaction coefficient $Z \neq 0$ being determined from a known function over the solutions of (\ref{res}).

System (\ref{ch4:complexB0}) possesses three conservation laws, first found  in \cite{MR56}  called
Manley--Rowe constants (MRC) of motion.
For a resonant triad, the MRC may be written as:
\be
I_{13} =  |B_1 |^2 + |B_3|^2 \,,\quad
I_{23} = |B_2 |^2 + |B_3|^2 \,, \ \ I_{12}= |B_1 |^2 - |B_2 |^2\,,\label{MR1}
\ee
They are linearly dependent, but any two of them form a linearly independent subset, enough to provide the integrability of (\ref{ch4:complexB0}), \cite{CUP}.

In a 3-wave system, a resonant triad is called a primary resonance cluster, and clusters consisting of more than 3 modes (common clusters) may be decomposed into triads having joint modes, \cite{K09a}. Common clusters describe many real physical phenomena: in nonlinear water wave systems \cite{CHS96}, in piezoelectric semiconductors \cite{MW67}, in geophysics \cite{DKPP06,KL07}, in optics \cite{New11}  etc.

In this paper we present a simple constructive method for deducing generalized MRC (gMRC) for a common resonance cluster.  The gMRC are polynomials on $|B_j|^2$ with integer coefficients. The method has been developed by EK for the course on nonlinear resonance analysis held at the J. Kepler University, Linz, since 2005. A \emph{Mathematica} code based on this method has been written by Loredana Tec and can be downloaded from \cite{progr}. The method has been used afterwards by the author, collaborators and students but its description has not been published. Recent publication \cite{HBN12} where gMRC are studied by far more complicated methods indicates that a simpler algorithm and available program code may be of interest for physicists working in the area of discrete wave turbulence. 

\section{Computation of gMRC}\label{MNR}

\subsection{Computation of MRC } Let us rewrite (\ref{ch4:complexB0}) as
\bea
\dot{B}_1=  Z B_2^*B_3,\,
\dot{B}_2=  Z B_1^* B_3, \, \dot{B}_3=  - Z B_1 B_2,\label{ch4:complexB1} \\
\dot{B}_1^*=  Z B_2B_3^*,\,
\dot{B}_2^*=  Z B_1 B_3^*, \, \dot{B}_3^*=  - Z B_1^* B_2^*,\label{ch4:complexB2}
\eea
where the second equation is the complex conjugate of the first. As $
{B}_j B_j^*=|{B}_j|^2$ and
\be {\textbf{d}}(B_j {B}^*_j)/{\textbf{d} t} = B_j {\textbf{d}}{B}^*_j /{\textbf{d} t}  +
B^*_j {\textbf{d}}{B}_j/{\textbf{d} t}  ,
\ee
multiplication of each equation for $B_j$ and $B_j^*$ by $B_j^*$ and  $B_j$, respectively
yields
\bea \label{3w-dyn-cc-1}
\begin{cases}
{\textbf{d}}|{B}_1|^2/{\textbf{d} t} =   Z B_1^*B_2^*B_3 + Z B_1B_2B_3^*\\
{\textbf{d}}|{B}_2|^2/{\textbf{d} t} =   Z B_1^* B_2^*B_3 +   Z B_1 B_2 B_3^*, \\
{\textbf{d}}|{B}_3|^2/{\textbf{d} t} = -  Z B_1 B_2B_3^* -  Z B^*_1 B^*_2B_3.
\end{cases}
\RA  \nonumber \\
\begin{cases}
{\textbf{d}}|{B}_1|^2/{\textbf{d} t} - {\textbf{d}}|{B}_2|^2/{\textbf{d} t} = 0,\\
{\textbf{d}}|{B}_1|^2/{\textbf{d} t} + {\textbf{d}}|{B}_3|^2/{\textbf{d} t} = 0, \\
{\textbf{d}}|{B}_2|^2/{\textbf{d} t} + {\textbf{d}}|{B}_3|^2/{\textbf{d} t} = 0,
\end{cases}
\RA   \begin{cases}
I_{12}=|B_1 |^2 -|B_2 |^2= \const,\\
I_{13}= |B_1 |^2 + |B_3|^2 = \const, \\
I_{23}=|B_2 |^2 + |B_3|^2 = \const,
\end{cases}
\eea
which is the set of MRC found originally in \cite{MR56}.
\subsection{Computation of gMRC for a two triad cluster}
Regard a simple common cluster in a three-wave system  formed by two triads $a$ and $b$ connected by the low frequency modes $B_{1a}=B_{1b}.$ The corresponding dynamical system reads
\bea \label{gen-but}
\begin{cases}
\dot{B}_{1a} =   Z_a B_{2a}^*B_{3a} +   Z_b B_{2b}^*B_{3b},\\
\dot{B}^*_{1a} =   Z_a B_{2a}B_{3a}^* +   Z_b B_{2b}B_{3b}^*,\\
\dot{B}_{2a} =   Z_a B_{1a}^* B_{3a}, \ \dot{B}^*_{2a} =   Z_a B_{1a} B_{3a}^*, \\
\dot{B}_{3a} =  -Z_a B_{1a} B_{2a}, \ \dot{B}^*_{3a} =  -Z_a B_{1a}^* B_{2a}^*, \\
\dot{B}_{2b}=  Z_b B_{1a}^* B_{3b}, \ \dot{B}^*_{2b}=  Z_b B_{1a}^* B_{3b}, \\
\dot{B}_{3b} =  -Z_b B_{1a} B_{2b},\dot{B}^*_{3b} =  -Z_b B^*_{1a} B^*_{2b},
\end{cases}
\eea
Considerations similar to those above yield three gMRC:
\bea
\begin{cases}
{\textbf{d}}|{B}_{1a}|^2/{\textbf{d} t}  =   Z_a B_{1a}^*B_{2a}^*B_{3a} +   Z_b B_{1a}^*B_{2b}^*B_{3b} \nn \\
 +  Z_a B_{1a}B_{2a}B_{3a}^* +   Z_b B_{1a}B_{2b}B_{3b}^*,\nn \\
{\textbf{d}}|{B}_{2a}|^2/{\textbf{d} t} =  Z_a B_{1a}^* B_{2a}^*B_{3a} +   Z_a B_{1a} B_{2a} B_{3a}^*, \nn \\
{\textbf{d}}|{B}_{3a}|^2/{\textbf{d} t}  = -  Z_a B_{1a} B_{2a}B_{3a}^* -  Z_a B^*_{1a} B^*_{2a}B_{3a}.\nn \\
{\textbf{d}}|{B}_{2b}|^2/{\textbf{d} t}  =   Z_b B_{1a}^* B_{2b}^*B_{3b} +   Z_b B_{1a} B_{2b} B_{3b}^*, \nn \\
{\textbf{d}}|{B}_{3b}|^2/{\textbf{d} t}  = -  Z_b B_{1a} B_{2b}B_{3b}^* -  Z_b B^*_{1a} B^*_{2b}B_{3b},\nn
\end{cases}
\RA \nonumber \\
\begin{cases}
{\textbf{d}}(|{B}_{1a}|^2+|{B}_{3a}|^2+|{B}_{3b}|^2) /{\textbf{d} t}=0,\\
{\textbf{d}}(|{B}_{2a}|^2+ |{B}_{3a}|^2) /{\textbf{d} t}=0,\\
{\textbf{d}}(|{B}_{2b}|^2+ |{B}_{3b}|^2) /{\textbf{d} t}=0
\end{cases}
\RA\begin{cases}
|{B}_{1a}|^2+|{B}_{3a}|^2+|{B}_{3b}|^2=\const,\\
|{B}_{2a}|^2+ |{B}_{3a}|^2=\const,\\
|{B}_{2b}|^2+ |{B}_{3b}|^2=\const.
\end{cases}
\label{MR-PP}
\eea
It is important to note that the choice of the connecting mode influences the form of the generalized MRCs. Indeed, for the  connection $B_{1a}=B_{1b}$ we got the set of constants (\ref{MR-PP}) whereas the connection $B_{1a}=B_{3b}$ yields a different set of MRCs:
\bea \label{MR-AP}
\begin{cases}
I_{23|a}= |B_{2|a} |^2 + |B_{3|a}|^2 \,, \\
I_{12|b}=|B_{1|b} |^2 - |B_{2|b}|^2 \,, \\
 I_{|a,b}=|B_{3|a}|^2 +|B_{1|b}|^2+|B_{3|b} |^2 \,.
 \end{cases}
\eea
The sets (\ref{MR-PP}) and (\ref{MR-AP}) are linearly independent and the corresponding dynamical systems are not equivalent.

As it was first shown in  \cite{KL08} the connections within a cluster of triads may be classified into \emph{three connection types} depending on whether a connection involves both or one or none of the high frequency modes $B_{3a}$, $B_{3b}$.
A high frequency mode is unstable due to the criterion of decay instability and is called \emph{active} or A-mode while low frequency modes $B_{1a}, B_{2a}, B_{1b}, B_{2b}$ are called passive or P-modes. Each connection within a common cluster can be either of AA-, AP- or PP-type (see \cite{CUP} for more details). In the NR-diagram representation they are shown as two bold, bold-dashed and dashed-dashed half edges correspondingly and each triad is shown as a triangle

\begin{figure}
\begin{center}
\includegraphics[width=3cm,height=0.53cm]{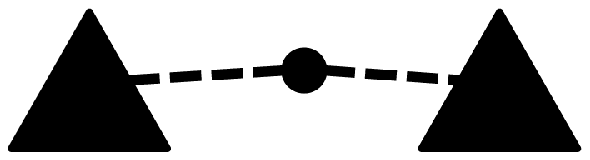}
\includegraphics[width=3cm,height=0.5cm]{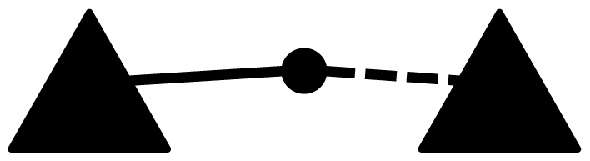}
\includegraphics[width=3cm,height=0.5cm]{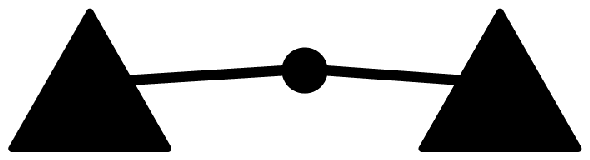}
\includegraphics[width=10cm,height=1cm]{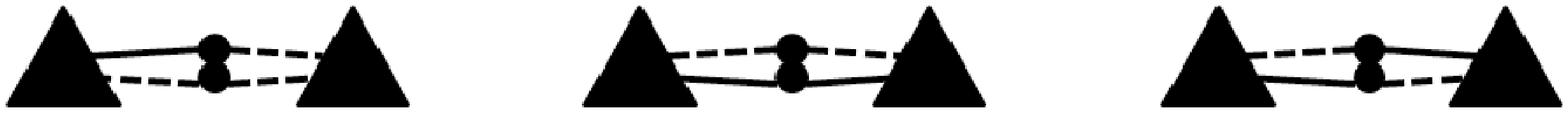}
\caption{\label{But} NR-diagrams for all 2-triad clusters}
\end{center}
\end{figure}

\begin{figure}[h]
\begin{center}\includegraphics[width=10cm,height=5cm]{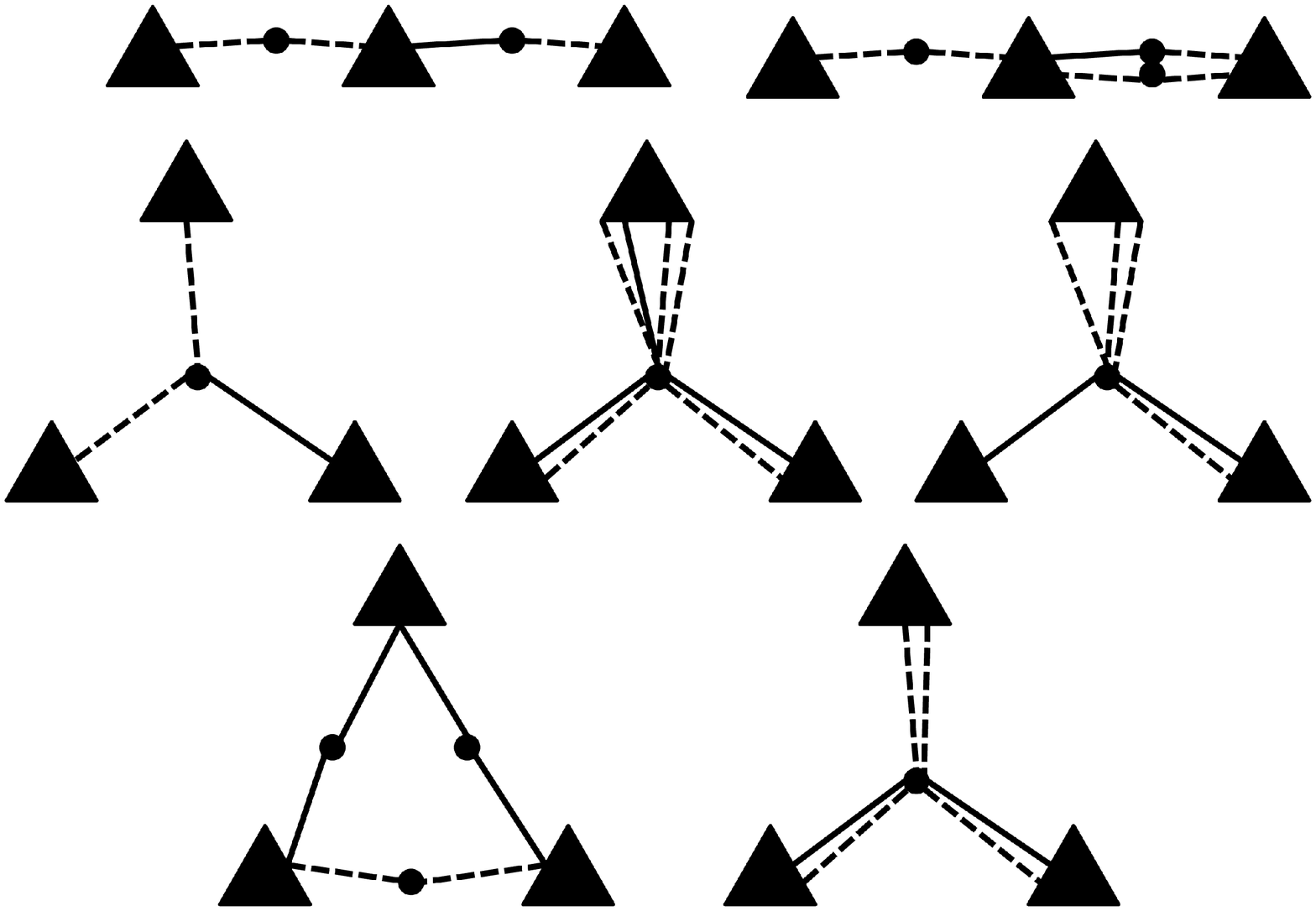}
\end{center}
\caption{\label{3tr-NR} NR-diagrams for some 3-triad clusters}
\end{figure}

Examples of all possible NR-diagrams for 2-triad clusters are shown in Fig.\ref{But}. For the NR-diagrams shown in the upper panel on the left and in the middle the sets of MRCs are given in (\ref{MR-PP}) and (\ref{MR-AP}) respectively.
The variety of 3-triad clusters is substantially richer (a few dozen). Examples of NR-diagrams for some possible clusters are shown in Fig.\ref{3tr-NR}. In real physical applications, clusters of 5 to 10 triads and more are common, e.g. \cite{KL08,KL07,all08,LPPR09}. The sheer voluminous of computations for deducing MRCs makes the use of a computer algebra system mandatory.

Another important issue is that
as any linear combination of gMRCs is a gMRC, gMRCs are not unique;   any of the constants in (\ref{MR-PP}) can be replaced by some of their linear combinations:
\bea
I_1=|{B}_{1a}|^2- |{B}_{2a}|^2 +|{B}_{3b}|^2, \mbox{  or  }\\
I_2=|{B}_{1a}|^2+|{B}_{2a}|^2 + 2|{B}_{3a}|^2+|{B}_{3b}|^2, \mbox{  or else.  }
\eea
However, for qualitative and numerical studies
we are naturally interested in computing "a minimal set" of gMRC, i.e a set of the shortest possible form which contains specific variables.
For instance,  for the dynamical system (\ref{gen-but}), the set
\bea \label{min1}
\begin{cases}
I_{23|a}=|{B}_{2a}|^2+ |{B}_{3a}|^2, \ \
I_{23|b}=|{B}_{2b}|^2+ |{B}_{3b}|^2,\\
I_{|ab}=|{B}_{1a}|^2+|{B}_{3a}|^2+|{B}_{3b}|^2
\end{cases}
\eea
is minimal while the set
\bea \label{non-min1}
\begin{cases}
I_{23|a}=|{B}_{2a}|^2+ |{B}_{3a}|^2, \ \
I_{23|b}=|{B}_{2b}|^2+ |{B}_{3b}|^2,\\
I_2=|{B}_{1a}|^2+|{B}_{2a}|^2 + 2|{B}_{3a}|^2+|{B}_{3b}|^2.
\end{cases}
\eea
is not. Minimal sets of gMRC can be computed using Gr\"obner bases.

\section{Implementation details}\label{method}

The computation of gMRC has been
implemented in \emph{Mathematica} as a package \texttt{NonlinearResonance} including
the following procedures (algorithms are given below):
\begin{itemize}
\item  \texttt{DynamicalSystem} for computing dynamical systems for common clusters in a three wave system;
\item \texttt{ConservationLawsLinComb} for computing the MRCs  for a given cluster by considering all linear combinations of the variables from the corresponding dynamical system having coefficients from $\{0,\pm 1\}$. The set of conservation laws is obtained by computing the null space of the set of these vertices;
\item \texttt{ConservationLawsGb} for computing a Gr\"obner basis of the ideal of the MRCs for a common cluster.
\end{itemize}
\begin{algorithm}
\caption{DynamicalSystem}\label{DynamicalSystem}
{\footnotesize
\begin{algorithmic}[1]
\REQUIRE $L=((L_{ij})_{j=a}^c)_{i=1}^s$ a list of triads representing a cluster.
\ENSURE The dynamical system $R$ of the cluster.
\STATE $M:=(Z_1B^\ast_{1b},B_{1c},Z_1B^\ast_{1a}B_{1c},-Z_1B_{1a}B_{1b},\ldots,Z_sB^\ast_{sb},B_{sc},Z_sB^\ast_{sa}B_{sc},-Z_sB_{sa}B_{sb})$;\\$N:=(B_{1a},B_{1b},B_{1c},\ldots,B_{sa},B_{sb},B_{sc})$;\\$\tilde L:=(L_{1a},L_{1b},L_{1c},\ldots,L_{sa},L_{sb},L_{sc})$;
\FOR{$i$ from $3s$ by $-1$ to $2$}
\FOR{$j$ from $i-1$ by $-1$ to $1$}
\IF{$\tilde L_i=\tilde L_j$}
\STATE $M_j:=M_i+M_j$;\\$M_i:=NULL$;\\In $M$ substitute $N_i$ by $N_j$;\\$N_i:=NULL$;\\$j:=1$;
\ENDIF
\ENDFOR
\ENDFOR
\FOR{$i$ from $1$ to $|M|$}
\STATE $R_{i}:=(\frac{d}{dt} N_i=M_i)$;\\$R_{i+|M|}:=(\frac{d}{dt} N_i^\ast=M_i^\ast)$;
\ENDFOR
\RETURN $R$;
\end{algorithmic}
}
\end{algorithm}
\begin{algorithm}
\caption{ConservationLawsLinComb}\label{ConservationLawsLinComb}
{\footnotesize
\begin{algorithmic}[1]
\REQUIRE $L$ a list of triads representing a cluster.
\ENSURE A minimal set $M$ of MRCs for the cluster.
\STATE $\tilde L:=$DynamicalSystem($L$);
\STATE Let $n$ be the number of nodes appearing in the dynamical system $\tilde L$, say $B_{\lambda_1},\ldots,B_{\lambda_n}$;
\FOR{$1\leq i\leq n$}
\STATE $\frac{d}{dt}(|B_{\lambda_i}|^2):=B_{\lambda_i}\frac{d}{dt}(B_{\lambda_i}^\ast)+\frac{d}{dt}(B_{\lambda_i})B_{\lambda_i}^\ast$, where $\frac{d}{dt}(B_{\lambda_i}),\frac{d}{dt}(B_{\lambda_i}^\ast)$ are determined by the dynamical system;
\ENDFOR
\STATE Determine a basis $M$ of$$\left\{(v_1,\ldots,v_n)\in\{0,\pm 1\}^n~\left|~\frac{d}{dt}\left(v_1|B_{\lambda_1}|^2+\ldots+v_n|B_{\lambda_n}|^2\right)=0\right.\right\};$$
\RETURN $M$;
\end{algorithmic}
}
\end{algorithm}

\begin{algorithm}
\caption{ConservationLawsGb}\label{ConservationLawsGb}
{\footnotesize
\begin{algorithmic}[1]
\REQUIRE $L$ a list of triads representing a cluster.
\ENSURE A Gr\"obner basis $G$ of the ideal of conservation laws and the corresponding term order $\prec$.
\STATE $\tilde L:=$DynamicalSystem($L$);
\STATE Let $n$ be the number of nodes appearing in the dynamical system $\tilde L$, say $B_{\lambda_1},\ldots,B_{\lambda_n}$;
\FOR{$1\leq i\leq n$}
\STATE $\frac{d}{dt}(|B_{\lambda_i}|^2):=B_{\lambda_i}\frac{d}{dt}(B_{\lambda_i}^\ast)+\frac{d}{dt}(B_{\lambda_i})B_{\lambda_i}^\ast$, where $\frac{d}{dt}(B_{\lambda_i}),\frac{d}{dt}(B_{\lambda_i}^\ast)$ are determined by the dynamical system;
\ENDFOR
\STATE Choose a term order $\prec$ such that for $t_1\in[\frac{d}{dt}(|B_{\lambda_1}|^2),\ldots,\frac{d}{dt}(|B_{\lambda_n}|^2)],~1\neq t_2\in[B_{\lambda_1},\ldots,B_{\lambda_n},B_{\lambda_1}^\ast,\ldots,B_{\lambda_n}^\ast],~t_3\in[\frac{d}{dt}(|B_{\lambda_1}|^2),\ldots,\frac{d}{dt}(|B_{\lambda_n}|^2),B_{\lambda_1},\ldots,B_{\lambda_n},B_{\lambda_1}^\ast,\ldots,B_{\lambda_n}^\ast]$ we have $t_1\prec t_2t_3$ and such that $\prec$ restricted to $[\frac{d}{dt}(|B_{\lambda_1}|^2),\ldots,\frac{d}{dt}(|B_{\lambda_n}|^2)]$ is degree lexicographic;
\STATE Compute a Gr\"obner basis $\tilde G$ of the ideal $\langle\frac{d}{dt}(|B_{\lambda_i}|^2)-B_{\lambda_i}\frac{d}{dt}(B_{\lambda_i}^\ast)+\frac{d}{dt}(B_{\lambda_i})B_{\lambda_i}^\ast~|~i=1,\ldots,n\rangle\trianglelefteq \mathbb Q[\frac{d}{dt}(|B_{\lambda_1}|^2),\ldots,\frac{d}{dt}(|B_{\lambda_n}|^2),B_{\lambda_1},\ldots,B_{\lambda_n},B_{\lambda_1}^\ast,\ldots,B_{\lambda_n}^\ast]$ with respect to $\prec$;
\STATE $G=\tilde G\cap\mathbb Q[\frac{d}{dt}(|B_{\lambda_1}|^2),\ldots,\frac{d}{dt}(|B_{\lambda_n}|^2)]$;
\STATE Remove all elements of $G$ with degree $\geq 2$;
\STATE In $G$ substitute $\frac{d}{dt}(|B_{\lambda_i}|^2)$ by $|B_{\lambda_i}|^2$ for $i=1,\ldots,n$;
\RETURN $G$ and $\prec$;
\end{algorithmic}
}
\end{algorithm}
Each of those takes as input a list of triads, where every triad is represented by a list of its vertices. The procedure \texttt{DynamicalSystem} returns the dynamical system of the cluster specified by the input. The procedure \texttt{ConservationLawsLinComb} returns a ''minimal`` basis of the set of conservation laws -- i.e., a set of MRCs -- corresponding to the input cluster. The procedure \texttt{ConservationLawsGb} returns a normalized reduced Gr\"obner basis of the ideal of conservation laws with respect to the lexicographic term ordering on the $|B_{ij}|^2$s such that$$|B_{ij}|^2>|B_{kl}|^2\begin{cases}\text{if }i<k\text{ or}\\\text{if }i=k \text{ and }j<l\end{cases}.$$This is usually computed faster than the basis returned by \texttt{conservationlaws} but its elements need not only have coefficients $\pm 1$ and $0$.
Computation time for arbitrary clusters consisting of 20-30 triads is about a few seconds.
\section{Discussion}

The construction of any constants of motion aims usually at establishing the integrability of the underlying dynamical system. For instance, integrability of a two triad cluster with $B_{3a}=B_{3b}$ and $Z_a/Z_b=2$
is proven in \cite{Ver68a} where an additional constant of motion (a polynomial of sixth order) has been constructed via direct search based on irreducible forms. On the other hand, a two triad cluster with $B_{1a}=B_{1b}$ and $Z_a/Z_b=3/4$ shows chaotic energy exchange among the modes of the cluster (see Poincar\'{e} sections for this case in \cite{CUP}, Fig.4.6).

Examples of integrable dynamical systems are known only for  resonance clusters of a very special form and/or for special initial conditions. Moreover,
there is no general method to establish integrability of a dynamical system and usually only a small set of constants of motion of any form (other than gMRC) can be found, e.g.
\cite{CUP,Ver68a,MCL83a,MCL83b,Ver68b,BK09_3}.

The algorithm described in this paper allows to produce in a matter of seconds (making use of a computer algebra system) a set of gMRC which is appropriate for qualitative study of the energy transport in a resonance cluster, for simplifying numerical study by reducing the number of variables in a dynamical system or for checking stability of the chosen numerical scheme.

Generalized MRCs can be analogously constructed for a 4-wave system where the primary clusters are quartets. The explicit form of the gMRCs for a quartet can be found in \cite{CUP}.

\section*{Acknowledgments}
E.K. acknowledges the support of the Austrian
Science Foundation (FWF) under project P22943-N18 "Nonlinear resonances of water waves" and in part -- by the Project of Knowledge Innovation Program (PKIP) of Chinese Academy of Sciences, Grant No. KJCX2. YW. W10. E.K. is very much obliged
to the organizing committee of the program "New Directions in Turbulence" (KITPC/ITP-CAS, 2012) and the
hospitality of Kavli ITP, Beijin, where part of this work has been accomplished.
L.T. is supported by the Austrian Academy of Sciences  as the recipient of a DOC-fFORTE-fellowship.


\end{document}